\documentclass[preprint,12pt]{elsarticle}
\usepackage{amssymb}
\usepackage{amsmath}

\begin{document}

\begin{frontmatter}

\title{Quantum public-key algorithms to encrypt and authenticate quantum messages with information-theoretic security}

\author{Min Liang}
\author{Li Yang\corref{1}}\ead{yangli@iie.ac.cn}
\cortext[1]{Corresponding author.}
\address{Institute of Information Engineering, Chinese Academy of Sciences, Beijing, China}
\address{Graduate University of Chinese Academy of Sciences, Beijing, China}

\begin{abstract}
Public-key cryptosystems for quantum messages are considered from two aspects: public-key encryption and public-key authentication. Firstly, we propose a general construction of quantum public-key encryption scheme, and then construct an information-theoretic secure instance. Then, we propose a quantum public-key authentication scheme, which can protect the integrity of quantum messages. This scheme can both encrypt and authenticate quantum messages. It is information-theoretic secure with regard to encryption, and the success probability of tampering decreases exponentially with the security parameter with regard to authentication. Compared with classical public-key cryptosystems, one private-key in our schemes corresponds to an exponential number of public-keys, and every quantum public-key used by the sender is an unknown quantum state to the sender.
\end{abstract}

\begin{keyword}
Quantum cryptography \sep authentication \sep quantum public-key \sep private quantum channel
\end{keyword}

\end{frontmatter}

\newcounter{definition}
\newcounter{theorem}
\newcounter{lemma}
\newcounter{proposition}
\newcounter{corollary}
\newcounter{remark}

\section{Introduction}
There are three kinds of cryptosystems for quantum messages, such as quantum no-key protocol \cite{yang02,yang11}, private quantum channel (or quantum one-time pad) \cite{boykin2000,boykin2002,ambainis2000}, and quantum public-key encryption protocol \cite{yang2003a}.

Quantum one-time pad \cite{boykin2000,boykin2002} was proposed to encrypt $n$ qubits using $2n$-bit secret key. Presharing $2n$-bit secret key is sufficient and necessary for encrypting $n$-qubit messages with perfect security. Ambainis et al. \cite{ambainis2000} defined private quantum channel (PQC), which is actually the same as quantum one-time pad. PQC (or quantum one-time pad) is a type of symmetric-key encryption scheme for quantum messages and it is considered to have perfect security. Later, others \cite{hayden2004,ambainis2004} relaxed the security requirement of PQC, and proposed approximate private quantum channel (APQC) (or approximate randomization of quantum state). This relaxation reduced the length of preshared classical key.

Leung \cite{leung2001} proposed another kind of quantum one-time pad with preshared EPR pairs as the secret key. In their scheme, the secret key can be reused securely. In addition, \cite{zhou2005,zhou2006} studied realizable quantum block encryption algorithm based on some simple bit-wise quantum computation. All these researches are quantum-message-oriented encryption schemes with preshared secret key.

Yang~\cite{yang2003a} constructed the first quantum-message-oriented public-key encryption protocol with classical private-
key and classical public-key. It is a computationally secure
quantum public-key encryption protocol.
Kawachi and Portmann \cite{kawachi2008} presented another kind of quantum-message-oriented public-key encryption protocol, where the public-key is the quantum state. By analyzing the protocol from the message size and the number of copies of the quantum public-key, they showed that it is bounded information-theoretic secure.

In this paper, we propose a quantum-message-oriented public-key encryption protocol, where one private-key corresponds to an exponential number of quantum public-keys and any two public-keys are different. In this scheme, the quantum public-keys are unknown to the sender, and the sender can only use them. The scheme has been proved to be truly information-theoretic secure.

Quantum authentication scheme (QAS) was firstly defined by Barnum et al. \cite{barnum2002}. They showed that any scheme to
authenticate quantum messages must also encrypt them, and constructed a quantum-message-oriented symmetric-key authentication scheme with preshared classical key. Their scheme can both encrypt and authenticate $n$-qubit message. If encrypting and authenticating $n$-qubit message into $n+q$ qubits, the sender and receiver need to preshare $2n+O(q)$-bit classical key, where $q$ is the security parameter.
Later, \cite{yang2003b,yang2010a} constructed a quantum-message-oriented public-key authentication scheme without a preshared classical key. However, its security is based on computational assumptions. Zhang \cite{zhang2009} proposed another type of QAS to authenticate the identity of the users, which will not be studied in this paper.

We propose a quantum-message-oriented public-key authentication scheme with the public-keys being quantum states. It can both encrypt and authenticate quantum messages. It is information-theoretic secure with regard to encryption, and the success probability of tampering decreases exponentially with the security parameter with regard to authentication.

\section{Preliminaries}\label{sec2}  

\subsection{Private quantum channel~\cite{ambainis2000}}  
Ambainis et al. \cite{ambainis2000} defined PQC with an ancillary quantum state. Here we use the PQC without ancillary qubits. The definition is as follows.

\stepcounter{definition}
{\bf Definition \arabic{definition}:} PQC is a set $\{p_i, U_i | i=1,2,...,2^{2n}\}$, where $\sum_{i=1}^{2^{2n}}p_i=1$, $p_i$ is the probability of using the classical key $i$. The PQC is used in the following way: Alice and Bob preshare a classical secret key $i$, then
\begin{enumerate}
\item{}
Alice uses the unitary transformation $U_i$ to encrypt a $n$-qubit message $\sigma$, and obtains its quantum cipher $\sigma'=U_i\sigma U_i^\dagger$. Alice sends $\sigma'$ to Bob.
\item{}
Bob uses the unitary transformation $U_i^\dagger$ to decrypt $\sigma'$, and obtains the message $\sigma=U_i^\dagger\sigma' U_i$.
\end{enumerate}
In order to be secure, it is required that the following formula holds for any $n$-qubit state $\sigma$:
\begin{equation}\label{eqn5}
\sum_{i\in\{0,1\}^{2n}}p_iU_i\sigma U_i^\dagger=\sigma_0,\forall\sigma,
\end{equation}
where $\sigma_0$ is a fixed state which is independent of $\sigma$ (For example $\sigma_0=\frac{1}{2^n}I$).

PQC is a symmetric-key cryptosystem using a preshared classical key. Denote $PQC_l(\sigma)$ as using $2n$-bit classical key $l\in\{0,1\}^{2n}$ to encrypt $n$-qubit message $\sigma$ through PQC, and its quantum cipher is denoted as $l^{(\sigma)}$. For example, \cite{boykin2000,boykin2002} proposed a PQC $\{p_{\alpha,\beta}=\frac{1}{2^{2n}},U_{\alpha,\beta}=X^\alpha Z^\beta|\alpha,\beta\in\{0,1\}^{n}\}$. Its encryption transformation is
\begin{eqnarray}\label{eqn1}
l^{(\sigma)}&=&PQC_l(\sigma)= U_l\sigma U_l^\dagger \nonumber\\
&=&\left(\otimes_{j=1}^n X^{l_j}\right)\cdot\left(\otimes_{j=n+1}^{2n}Z^{l_j}\right)\sigma\left(\otimes_{j=n+1}^{2n}Z^{l_j}\right)\cdot\left(\otimes_{j=1}^n X^{l_j}\right).
\end{eqnarray}
The PQC decryption transformation is
\begin{eqnarray}
&& PQC_l^{-1}(\sigma)= U_l^\dagger\sigma U_l \nonumber\\
&=&\left(\otimes_{j=n+1}^{2n}Z^{l_j}\right)\cdot\left(\otimes_{j=1}^n X^{l_j}\right)\sigma\left(\otimes_{j=1}^n X^{l_j}\right)\cdot\left(\otimes_{j=n+1}^{2n}Z^{l_j}\right).
\end{eqnarray}

Other researchers \cite{hayden2004,ambainis2004} studied the approximate quantum encryption or approximate PQC (APQC). In APQC, the security condition Eq.(\ref{eqn5}) is relaxed in order to lessen the preshared secret key. It is required that
\begin{equation}
D\left(\sum_{i\in\{0,1\}^m}p_iU_i\sigma U_i^\dagger,\sigma_0\right)\leq \epsilon,
\end{equation}
where $\sigma$ is any $n$-qubit message, $m$ is the length of the preshared key, and $m<2n$. $D(\rho_1,\rho_2)=\frac{1}{2}tr|\rho_1-\rho_2|$ represents the trace distance of two density matrixes $\rho_1$ and $\rho_2$ \cite{nielsen2000}. $\epsilon$ is security parameter, and APQC is considered a perfect PQC when $\epsilon=0$.

We denote $APQC_l(\sigma)$ as using $m$-bit preshared key $l\in\{0,1\}^m$ to encrypt $n$-qubit message $\sigma$ through APQC. For example, we can adopt the last scheme (hybrid construction) in \cite{ambainis2004}. That scheme is described as follows.
Let $B$ be a $\delta$-biased set on $n$ bits. For $b\in B$, define a unitary transformation $U_b$ as follows. Define $U_b|x\rangle =(-1)^{b\cdot x}|x\rangle, \forall x\in\{0,1\}^n$, where $b\cdot x$ is
the usual (bitwise) inner product of $b$ and $x$. Each $a\in\{0,1\}^n$ and $b\in B$ are selected with uniform probability. The APQC transformation can be described as follows:
\begin{equation}
APQC(\sigma)=\frac{1}{2^n}\cdot\frac{1}{|B|}\sum_{a\in\{0,1\}^n}\sum_{b\in B}APQC_{a,b}(\sigma),
\end{equation}
where $APQC_{a,b}(\sigma)$ represents using the preshared classical key $a||b$ to encrypt quantum message $\sigma$ ("$||$" denotes an concatenation of two bit-strings).
\begin{equation}\label{eqn6}
APQC_{a,b}(\sigma)=X^a Z^{a^2}U_b\sigma U_b^\dagger Z^{-a^2}X^{-a},
\end{equation}
where $X^a=X^{a_1}\otimes\cdots\otimes X^{a_n}$, $X^{-a}=(X^a)^\dagger$ and $a\in\{0,1\}^n,b\in B\subset\{0,1\}^n$. The total length of $a$ and $b$ is $n+n=2n$. Because $B$ is a $\delta$-biased set on $n$ bits, $n'=log|B|+O(1)$ bits of randomness is enough to generate any $n$-bit number $b\in B$ in polynomial time \cite{ambainis2004}. In other words, the set $B$ can be generated from the set $\{0,1\}^{n'}$ using a polynomial-time algorithm. For convenience, each number $b\in B$ can be seen as one element of the set $\{0,1\}^{n'}$.
Thus, this APQC construction needs only $m=n+n'$ bits of the classical key, and $m=n+log|B|+O(1)<2n$.

\subsection{Quantum authentication scheme~\cite{barnum2002}}
Authentication of quantum messages was defined by Barnum et al. \cite{barnum2002}. A sender Alice and a receiver Bob must preshare a classical key $k\in K$.
Alice and Bob use $k$ to authenticate the quantum message.

\stepcounter{definition}
{\bf Definition \arabic{definition}:} QAS is defined by a triplet $(A,D,K)$, where $A$ and $D$ are two polynomial-time quantum algorithms, and $K$ is a set of classical keys. $(A,D,K)$ satisfies:
\begin{enumerate}
\item{}
Alice performs quantum algorithm $A$ on a $n$-qubit message $\sigma$ and a classical key $k\in K$, and outputs a $n+t$-qubit state $\sigma'$. Alice sends $\sigma'$ to Bob.
\item{}
Bob receives a quantum state $\sigma'$, and then inputs $\sigma'$ and the classical key $k\in K$ to quantum algorithm $D$. The output of $D$ has two parts: a $n$-qubit message $\sigma$, a single-qubit $|v\rangle$. Bob decides to accept or reject according to the single-qubit $|v\rangle$ (accept if it is $|1\rangle$ and reject if it is $|0\rangle$).
\end{enumerate}

From this definition, QAS is a type of symmetric-key authentication for quantum messages. However, we will consider public-key authentication of quantum messages in this paper.

\section{Security notion}\label{sec3}
In this section, information-theoretic security is defined for public-key encryption of quantum messages, and two sufficient conditions are presented here.

\stepcounter{lemma}
{\bf Lemma \arabic{lemma}:} $H_C$ is a quantum state space. The following two statements are equivalent:

(1) There exists a fixed quantum state $\tau$, such that $D(\rho,\tau)\leq \epsilon, \forall \rho\in H_C$.

(2) $D(\rho_1,\rho_2)\leq \epsilon, \forall \rho_1,\rho_2\in H_C$.

{\bf Proof:} Firstly, the statement (2) can be deduced from (1). $\forall \rho_1,\rho_2\in H_C$,
\begin{eqnarray*}
D(\rho_1,\rho_2)&=&\frac{1}{2} tr|\rho_1-\rho_2|\\
&=& \frac{1}{2} tr|\rho_1-\tau+\tau-\rho_2|\\
&\leq & \frac{1}{2} tr|\rho_1-\tau|+\frac{1}{2} tr|\tau-\rho_2|\\
&=& D(\rho_1,\tau)+D(\rho_2,\tau)\\
&\leq & \epsilon+\epsilon=2\epsilon.
\end{eqnarray*}

It is straightforward to deduce (1) from (2). By randomly selecting a fixed quantum state $\tau$ from $H_C$, then $\tau$ can satisfy the condition $D(\rho,\tau)\leq \epsilon, \forall \rho\in H_C$.
$\hfill{}\Box$

From Definition 5.2.4 in \cite{goldreich2004}, indistinguishability was defined for public-key encryption of the classical messages.

\stepcounter{definition}
{\bf Definition \arabic{definition}:} A public-key encryption scheme for the classical messages has indistinguishable encryptions, if for every classical polynomial-size circuit family $\{C_n\}$, and every positive polynomial $p(\cdot)$, all sufficiently
large $n$, and every $x,y\in\{0,1\}^{poly(n)}$(i.e.,$|x|=|y|=n$),
\begin{equation}
\left|Pr[C_n(G(1^n),E_{G(1^n)}(x))=1]-Pr[C_n(G(1^n),E_{G(1^n)}(y))=1]\right| < \frac{1}{p(n)},
\end{equation}
where the algorithm $E$ is a classical encryption algorithm and $G$ is a algorithm for key generation.

From the discussion in Chapter 5.5.2 in \cite{goldreich2004}, the security can be classified according to the size of classical circuit family $\{C_n\}$: (1) if $\{C_n\}$ is polynomial-size, the above definition defines computational security; (2) if there are no limitations on the size of $\{C_n\}$, the above definition defines information-theoretic security.

Definition 3 in \cite{yang2010b} defines information-theoretic security of quantum public-key encryption for classical messages. It naturally extends the information-theoretic security of classical public-key encryption. It coincides with the notion of information-theoretically indistinguishable as discussed by Hayashi et al. \cite{hayashi2008}. Here, it is extended to information-theoretic security of quantum public-key encryption for quantum messages. Two sufficient conditions are presented here.

\stepcounter{definition}
{\bf Definition \arabic{definition}:} A quantum public-key encryption scheme for quantum messages is information-theoretic secure, if for every quantum circuit family $\{C_n\}$, every positive polynomial $p(.)$, all sufficiently large $n$, and any two quantum messages $\sigma,\sigma'\in H_M$, it holds that
\begin{equation}\label{eqn12}
\left|Pr[C_n(G(1^n),E_{G(1^n)}(\sigma))=1]-Pr[C_n(G(1^n),E_{G(1^n)}(\sigma'))=1]\right| <\frac{1}{p(n)},
\end{equation}
where the algorithm $E$ is a quantum algorithm for encryption and $G$ is a quantum algorithm for generating public-keys.

It should be noted that Yang et al. \cite{yang2010b} and in this paper, information-theoretic security are all defined using quantum circuit family $\{C_n\}$ without limitations on its size. This means $\{C_n\}$ can be any quantum circuit family of arbitrary size.
Here, the right side of Eq.(\ref{eqn12}) is $\frac{1}{p(n)}$, but it does not mean that the ciphers can be distinguished efficiently, because $p(n)$ is not a particular polynomial but an arbitrary polynomial. Thus the above definition means, for any two quantum messages, their quantum ciphers cannot be distinguished by any quantum circuit family of any size.

Next, two sufficient conditions are presented. The sender Alice encrypts a quantum message $\sigma\in H_M$ using a quantum public-key $\rho_k,k\in\mathcal{K}$. Its quantum cipher is denoted as $\rho_k^{(\sigma)}$.
Suppose each quantum public-key $\rho_k,k\in\mathcal{K}$ is used with probability $p_k$, and $\sum_{k\in\mathcal{K}}p_k=1$. The attacker Eve does not know the public-key used by Alice, so the quantum cipher (with respect to Eve) of $\sigma$ can be represented by $\sum_k p_k \rho_k^{(\sigma)}$.

\stepcounter{theorem}
{\bf Theorem \arabic{theorem}:} A quantum public-key encryption scheme $(E,G)$ for quantum messages is information-theoretic secure, if
for every positive polynomial $p(.)$, all sufficiently large $n$, any two quantum messages $\sigma,\sigma'\in H_M$,
\begin{equation}
D\left(\sum_k p_k \rho_k^{(\sigma)}, \sum_k p_k \rho_k^{(\sigma')}\right)<\frac{1}{p(n)},
\end{equation}
where $\rho_k^{(\sigma)}$ and $\rho_k^{(\sigma')}$ are quantum ciphers of $\sigma$ and $\sigma'$ using quantum encryption algorithm $E$ and public-key $\rho_k$, respectively.
We consider $p_k$ as the probability of generating public-key $\rho_k$ from the quantum algorithm $G$, and $\sum_{k\in\mathcal{K}}p_k=1$.

{\bf Proof:}
In Definition 4, $E$ is a quantum encryption algorithm which performs on quantum message $\sigma$, and $G$ is a quantum algorithm for generating public-keys, and each public-key $\rho_k$ is generated with a probability $p_k$, so
\begin{eqnarray*}
 Pr\left[C_n\left(G(1^n),E_{G(1^n)}(\sigma)\right)=1\right] &=& \sum_k p_k Pr\left[C_n\left(\rho_k^{(\sigma)}\otimes\sigma_a\right)=1\right]\\
&=& Pr\left[C_n\left(\sum_k p_k \rho_k^{(\sigma)}\otimes\sigma_a\right)=1\right],
\end{eqnarray*}
where $\sigma_a$ is a quantum state which acts as ancillary input to quantum circuit $C_n$.

Similarly, the following formula can be deduced.
\begin{equation*}
Pr\left[C_n\left(G(1^n),E_{G(1^n)}(\sigma')\right)=1\right] = Pr[C_n(\sum_k p_k \rho_k^{(\sigma')}\otimes\sigma_a)=1].
\end{equation*}

Any quantum circuit family $\{C_n\}$ built for distinguishing two density operators
corresponds to a set of positive operator-values measure (POVM) $\{E_m\}$.
We define $p_m = tr\left(C_n\left(\sum_k p_k \rho_k^{(\sigma)}\otimes\sigma_a\right)E_m\right)$ and \\ $q_m = tr\left(C_n\left(\sum_k p_k \rho_k^{(\sigma')}\otimes\sigma_a\right)E_m\right)$ as the probabilities of
measurement result labeled by $m$. In this case, we have
\begin{eqnarray}
&&\left|Pr\left[C_n\left(\sum_k p_k \rho_k^{(\sigma)}\otimes\sigma_a\right)=1\right]-Pr\left[C_n\left(\sum_k p_k \rho_k^{(\sigma')}\otimes\sigma_a\right)=1\right]\right|\nonumber\\
&\leq & \max\limits_{\{E_m\}}\frac{1}{2}\sum_m\left|tr\left[\left(C_n\left(\sum_k p_k \rho_k^{(\sigma)}\otimes\sigma_a\right)-C_n\left(\sum_k p_k \rho_k^{(\sigma')}\otimes\sigma_a\right)\right)E_m\right]\right|\nonumber\\
&=& \max\limits_{\{E_m\}} D(p_m,q_m)\label{eqn13}.
\end{eqnarray}
The formula Eq.(\ref{eqn13}) is equal to
\begin{eqnarray}
&& D\left(C_n\left(\sum_k p_k \rho_k^{(\sigma)}\otimes\sigma_a\right),C_n\left(\sum_k p_k \rho_k^{(\sigma')}\otimes\sigma_a\right)\right)\nonumber\\
&\leq & D\left(\sum_k p_k \rho_k^{(\sigma)}\otimes\sigma_a,\sum_k p_k \rho_k^{(\sigma')}\otimes\sigma_a\right)\nonumber\\
&=& D\left(\sum_k p_k \rho_k^{(\sigma)},\sum_k p_k \rho_k^{(\sigma')}\right)<\frac{1}{p(n)}.
\end{eqnarray}
Then,
\begin{equation*}
\left|Pr[C_n(G(1^n),E_{G(1^n)}(\sigma))=1]-Pr[C_n(G(1^n),E_{G(1^n)}(\sigma'))=1]\right|<\frac{1}{p(n)}.
\end{equation*}
Thus, the quantum public-key encryption scheme $(E,G)$ is information-theoretic secure. $\hfill{}\Box$

From Lemma 1 and Theorem 1, the following corollary can be deduced directly.

\stepcounter{corollary}
{\bf Corollary \arabic{corollary}:} A quantum public-key encryption scheme $(E,G)$ for quantum messages is information-theoretic secure, if
for every positive polynomial $p(.)$, all sufficiently large $n$, there exists a fixed quantum states $\tau$ such that
\begin{equation}
D\left(\sum_k p_k \rho_k^{(\sigma)},\tau\right)<\frac{1}{p(n)},\forall\sigma\in H_M,
\end{equation}
where $\rho_k^{(\sigma)}$ is the cipher of quantum message $\sigma$ using the quantum encryption algorithm $E$ and the public-key $\rho_k$.
Again, we consider $p_k$ as the probability of generating public-key $\rho_k$ from the quantum algorithm $G$, and $\sum_{k\in\mathcal{K}}p_k=1$. 

\section{Public-key encryption of quantum information}\label{sec4}
\subsection{A general construction}\label{subsec41}
Firstly, we define a model for quantum public-key encryption of classical messages \cite{pan2010}.

\stepcounter{definition}
{\bf Definition \arabic{definition}:} A public-key cryptosystem using quantum public-key to encrypt classical messages is described by a quadruple \cite{pan2010}: $$\Delta=\left(\{F_i\}_{i\in\mathcal{I}},\{(s,\rho_k)\}_{s\in\{0,1\}^{O(n)}},\mathcal{E},\mathcal{D}\right),$$
where all components are defined as follows.
\begin{itemize}
\item[1.] $\{F_i\}_{i\in\mathcal{I}}$ is a set of private-keys. Each $F_i$ is a polynomial-time computable function with $O(n)$-bit input and $n$-bit output. ($F_i:\{0,1\}^{O(n)}\rightarrow\{0,1\}^n$).
\item[2.] $\{(s,\rho_k)\}_{s\in\{0,1\}^{O(n)}}$ is a set of quantum public-keys. Each pair of $(s,k)$ is generated from a function $F\in\{F_i\}_{i\in\mathcal{I}}$, and their relation is $k=F(s)$. Given $k$, quantum state $\rho_k$ can be efficiently prepared.
\item[3.] $\mathcal{E}$ is a quantum encryption transformation. Alice uses $\mathcal{E}$ and quantum public-key $(s,\rho_k)$ to encrypt the classical message $l$, and obtains a quantum cipher $\rho_k^{(l)}=\mathcal{E}(\rho_k,l)$. Alice sends $(s,\rho_k^{(l)})$ to Bob.
\item[4.] $\mathcal{D}$ is a quantum decryption transformation. After receiving $(s,\rho_k^{(l)})$ from Alice, Bob computes $k=F(s)$ by using private-key $F\in\{F_i\}$. Bob then uses $\mathcal{D}$ and $k$ to decrypt the cipher $\rho_k^{(l)}$, and obtains the classical message $l=\mathcal{D}(\rho_k^{(l)},k)$.
\end{itemize}

In this definition, the private-key $F$ is a function which can be computed efficiently. From the key $F$, many different pairs of $(s,k)$ can be generated such that $F(s)=k$, thereby allowing many different quantum public-keys to be prepared. Thus, the relation between private-keys and public-keys is one-to-many. That means, a private-key $F$ corresponds to many quantum public-keys $(s,\rho_k)$, where $F(s)=k$.

In the encryption schemes introduced in \cite{kawachi2005,kawachi2008}, the relation between private-keys and public-keys is one-to-one (One private-key $k$ corresponds to one quantum public-key $\rho_k$). This kind of schema is a special case of the public-key cryptosystem $\Delta$. The reason is as follows. In $\Delta$, let $s\equiv 0$, then $k\equiv F(0)$, so $k$ is uniquely determined by $F$. Then the public-key $(0,\rho_k)$ is uniquely determined by private-key $F$.

\stepcounter{remark}
{\bf Remark \arabic{remark}:} Because the attacker Eve does not know the function $F$, she does not know the value of $k$. Suppose each $k$ is used with probability $p_k$, then the cipher of classical message $l$ (with respect to Eve) is $\sum_k p_k \rho_k^{(l)}$.

PQC is a symmetric-key encryption scheme for quantum messages with classical secret key.
$\Delta$ is a public-key encryption scheme for classical messages with quantum public-keys.
The two schemes are combined as Figure 1, and form a public-key encryption scheme for quantum messages with quantum public-keys.

\begin{figure}[htb!]
\begin{center}
\includegraphics[scale=0.7]{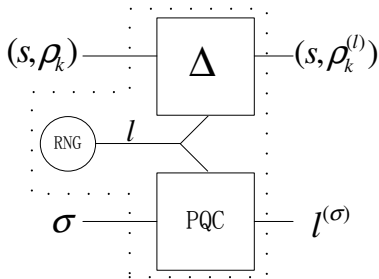}
\end{center}
\vspace{-5mm}
\caption{\label{fig1} Public-key encryption scheme for quantum messages with quantum public-keys.
RNG is a random number generator. $\Delta$ is a public-key encryption scheme for classical messages with quantum public-keys.
PQC is a symmetric-key encryption scheme for quantum messages with classical secret key.
RNG generates a random number $l$, then $l$ is used as an encryption key in PQC, and acts as a
classical message in $\Delta$. $(s,\rho_k)$ is a quantum public-key in $\Delta$.
Quantum state $\sigma$ is a message being encrypted by the PQC.}
\end{figure}

Next, we describe the scheme in detail.
Bob generates a private-key $F$ and many quantum public-keys, then he sends all quantum public-keys to his public-key register. The progress is as follows.

{\bf [Key Generation]}
\begin{enumerate}
\item{}
Bob randomly selects a function $F\in\{F_i\}_{i\in\mathcal{I}}$;
\item{}
Bob randomly selects some $s_j\in\{0,1\}^{O(n)}$, and computes $k_j=F(s_j)$, and then prepares quantum state $\rho_{k_j}$ according to $k_j$;
\item{}
Bob uploads all $(s_j,\rho_{k_j})$ to his public-key register. The function $F$ is Bob's private-key.
\end{enumerate}

Alice intends to send a $n$-qubit message $\sigma$ to Bob. She firstly downloads quantum public-keys from Bob's public-key register, then encrypts $\sigma$ with quantum public-keys. The progress is as follows.

{\bf [Encryption]}
\begin{enumerate}
\item{}
Alice randomly selects a $2n$-bit number $l$, and then encrypts the quantum message $\sigma$ by performing PQC encryption transformation $l^{(\sigma)}=PQC_l(\sigma)$;
\item{}
Alice downloads quantum public-key $(s,\rho_k)\in\{(s_j,\rho_{k_j})\}_j$ from Bob's public-key register, and then encrypts $l$ by performing encryption transformation $\mathcal{E}$, and obtains $\rho_{k}^{(l)}=\mathcal{E}(\rho_{k},l)$;
\item{}
Alice sends $s$ and the quantum cipher $\rho_{k}^{(l)}\otimes l^{(\sigma)}$ to Bob.
\end{enumerate}

Bob receives the cipher $(s,\rho_{k}^{(l)}\otimes l^{(\sigma)})$, and then decrypts it as follows.

{\bf [Decryption]}
\begin{enumerate}
\item{}
According to the value of $s$, Bob uses his private-key $F$ to compute $k=F(s)$;
\item{}
According to the value of $k$, Bob performs decryption transformation $\mathcal{D}$ on quantum cipher $\rho_{k}^{(l)}$, and obtains $l=\mathcal{D}(\rho_{k}^{(l)},k)$;
\item{}
According to the value of $l$, Bob performs PQC decryption transformation on quantum cipher $l^{(\sigma)}$, and obtains the quantum message $\sigma=PQC_l^{-1}(l^{(\sigma)})$.
\end{enumerate}

In this scheme, $l$ is a random number selected by Alice, and is unknown by Eve. From Remark 1, the cipher of the quantum message $\sigma$ (with respect to Eve) can be represented as follows.
\begin{equation}
\mathcal{G}(\sigma)=\frac{1}{2^{2n}}\sum_{l\in\{0,1\}^{2n}}\left[\sum_k p_k \rho_k^{(l)}\right]\otimes l^{(\sigma)}.
\end{equation}

This scheme is constructed by combining a public-key encryption scheme $\Delta$ and a symmetric-key encryption scheme PQC (or APQC).
The security of the combined scheme can be determined by the security of $\Delta$ and PQC (or APQC). The analysis is as follows.

\stepcounter{lemma}
{\bf Lemma \arabic{lemma}:} In a quantum state space $H_C$, if there exists a fixed state $\tau$, such that $D(\rho,\tau)\leq \epsilon, \forall \rho \in H_C$, then $D\left(\sum_{k\in\mathcal{K}} p_k \rho_k,\tau\right)\leq \epsilon$, where $\rho_k\in H_C$ and $\sum_{k\in\mathcal{K}} p_k=1$.

{\bf Proof:} From $\sum_{k\in\mathcal{K}}p_k=1$, it can be deduced that
\begin{eqnarray*}D\left(\sum_k p_k \rho_k,\tau\right) &=& \frac{1}{2}tr\left|\sum_k p_k \rho_k - \tau\right| = \frac{1}{2}tr\left|\sum_k p_k (\rho_k - \tau)\right|\\
&\leq & \frac{1}{2}tr\sum_k p_k \left|\rho_k - \tau\right|=\sum_k p_k D(\rho_k,\tau) \\
&\leq & \sum_k p_k \epsilon =\epsilon.\hfill{~~~~~~~~~~~~~~~~~~~~~~~~~~~~~~~~~~~~~~~~~}\Box
\end{eqnarray*}

\stepcounter{theorem}
{\bf Theorem \arabic{theorem}:} If there exists a scheme $\Delta$ and a PQC (APQC), which satisfy the following conditions:
\begin{enumerate}
\item{}
For the scheme $\Delta$, there exist a fixed state $\tau_1$, such that \\ $D\left(\sum\limits_k p_k \rho_k^{(l)},\tau_1\right)\leq\epsilon_1,\forall l$;
\item{}
For the PQC (APQC), there exist a fixed state $\tau_2$, such that \\ $D\left(\frac{1}{2^{2n}}\sum_{l\in\{0,1\}^{2n}} l^{(\sigma)},\tau_2\right)\leq\epsilon_2,\forall \sigma$;
\end{enumerate}
then there exists a fixed state $\tau$, such that $D\left(\mathcal{G}(\sigma),\tau\right)\leq\epsilon_1+\epsilon_2, \forall \sigma$.

{\bf Proof:} From the condition 1, $\forall l,\sigma$
\begin{equation*}
D\left(\left[\sum_k p_k \rho_k^{(l)}\right]\otimes l^{(\sigma)},\tau_1\otimes l^{(\sigma)}\right)=D\left(\sum_k p_k \rho_k^{(l)},\tau_1\right)\leq\epsilon_1.
\end{equation*}
According to Lemma 2, it can be deduced that
\begin{eqnarray}
D\left(\mathcal{G}(\sigma),\frac{1}{2^{2n}}\sum_{l\in\{0,1\}^{2n}}\tau_1\otimes l^{(\sigma)}\right)\leq\epsilon_1, \forall \sigma.
\end{eqnarray}
Then there exists a state $\tau=\tau_1\otimes\tau_2$, such that
\begin{eqnarray*}
&&D\left(\mathcal{G}(\sigma),\tau\right)=D\left(\mathcal{G}(\sigma),\tau_1\otimes \tau_2\right) \nonumber\\
& \leq & D\left(\mathcal{G}(\sigma),\frac{1}{2^{2n}}\sum_l\tau_1\otimes l^{(\sigma)}\right)+D\left(\frac{1}{2^{2n}}\sum_l\tau_1\otimes l^{(\sigma)},\tau_1\otimes \tau_2\right) \nonumber\\
& \leq &\epsilon_1+D\left(\frac{1}{2^{2n}}\sum_l l^{(\sigma)},\tau_2\right) \nonumber\\
& \leq &\epsilon_1+\epsilon_2.\hfill{~~~~~~~~~~~~~~~~~~~~~~~~~~~~~~~~~~~~~~~~~~~~~~~~~~~~~~~~~~~~~~~~~~~~~~~~}\Box
\end{eqnarray*}


\subsection{Quantum public-key encryption of classical information\cite{pan2010}}\label{subsec42}
In order to give a concrete example for the scheme described in Figure~\ref{fig1}, we firstly introduce an example for public-key encryption scheme $\Delta$. This example was proposed by Pan and Yang \cite{pan2010}. Their scheme can be used to encrypt single-bit classical message with information-theoretic security. Moreover, Yang et al. \cite{yang2011} proposed a classical message oriented quantum public-key scheme based on conjugate coding. This scheme is another example of $\Delta$.

Denote $\Omega_n=\{k\in\{0,1\}^n|there~has~odd~number~of~'1'~in~k\}$.
Define a $n$-qubit state $\rho_{k,i}=\frac{1}{2}(|i\rangle+|i\oplus k\rangle)(\langle i|+\langle i\oplus k|)$, where $i\in\{0,1\}^n$, $k\in\Omega_n$. Let $V_l=(\otimes_{j=1}^n Z)^{l},l\in\{0,1\}$, then it is the identity transformation $I$ while $l=0$, and it is the unitary transformation $\otimes_{j=1}^n Z$ while $l=1$. We define a transformation as follows:
\begin{equation}
\mathcal{T}:l\rightarrow V_l\rho_{k,i}V_l^\dagger, ~l\in\{0,1\}.
\end{equation}
Denote $\rho_{k,i}^{(l)}=\mathcal{T}(l)= V_l\rho_{k,i}V_l^\dagger$, then $\rho_{k,i}^{(0)}=\rho_{k,i}$. Because $k\in\Omega_n$, it can be deduced that $\rho_{k,i}^{(1)}=(\otimes_{j=1}^n Z)\rho_{k,i}(\otimes_{j=1}^n Z)^\dagger=\frac{1}{2}(|i\rangle-|i\oplus k\rangle)(\langle i|-\langle i\oplus k|)$.

According to Lemma 3 and Lemma 4 in \cite{pan2010}, if the values of $k$ and $i$ are unknown, it is information-theoretically indistinguishable between \\ $\frac{1}{2^{2n-1}}\sum\limits_{k\in\Omega_n}\sum\limits_{i\in\{0,1\}^n}\rho_{k,i}^{(0)}$ and $\frac{1}{2^{2n-1}}\sum\limits_{k\in\Omega_n}\sum\limits_{i\in\{0,1\}^n}\rho_{k,i}^{(1)}$ (the trace distance of them is $\frac{1}{2^{n-1}}$).
However, given the value of $k$, there exists a polynomial-time quantum algorithm which can distinguish $\frac{1}{2^n}\sum\limits_{i\in\{0,1\}^n}\rho_{k,i}^{(0)}$ and $\frac{1}{2^n}\sum\limits_{i\in\{0,1\}^n}\rho_{k,i}^{(1)}$. Therefore, $\mathcal{T}$ is a quantum trapdoor one-way transformation with trapdoor $k$.

The key generation process is as follows.
\begin{enumerate}
\item{}
Bob randomly selects an efficiently computable function $F:\{0,1\}^{O(n)}\rightarrow\{0,1\}^{n}$;
\item{}
Bob randomly selects a number $s\in\{0,1\}^{O(n)}$, and then computes a $n$-bit number $k=F(s)$. Then he continues the next step if $k$ is an element of $\Omega_n$, otherwise he randomly selects a new number $s$;
\item{}
Bob randomly selects a number $i\in\{0,1\}^n$, and prepares a $n$-qubit state $\rho_{k,i}=\frac{1}{2}(|i\rangle+|i\oplus k\rangle)(\langle i|+\langle i\oplus k|)$;
\item{}
Bob's public-key is $(s,\rho_{k,i})$, and private-key is $F$.
\end{enumerate}

Alice uses Bob's public-key to encrypt one classical bit $l\in\{0,1\}$. Its encryption transformation is as follows.
\begin{equation}\rho_{k,i}^{(l)}=\mathcal{E}(\rho_{k,i},l)=V_{l}\rho_{k,i}V_{l}^\dagger.\end{equation}
That means, Alice performs a unitary transformation $V_{l}=(\otimes_{j=1}^n Z)^{l}$ on Bob's public-key $(s,\rho_{k,i})$, and then sends its result $(s,\rho_{k,i}^{(l)})$ to Bob.

After receiving the cipher $(s,\rho_{k,i}^{(l)})$, Bob uses his private-key $F$ to decrypts as follows. He firstly computes $k=F(s)$ from the value of $s$, and then decrypts $\rho_{k,i}^{(l)}$ with its trapdoor $k$, and obtains the classical message $l$.

\stepcounter{remark}
{\bf Remark \arabic{remark}:} In this example, the public-key is $(s,\rho_{k,i})$. Compared with the encryption scheme $\Delta$ defined in Section~\ref{subsec41}, here a random parameter $i$ is added into the public-key in order to protect $k$ (See the analysis in Section~\ref{subsec44}).

\subsection{Quantum public-key encryption of quantum information}\label{subsec43}
According to the general construction in Section~\ref{subsec41}, and the example of $\Delta$ as introduced in Section~\ref{subsec42}, we can construct a concrete public-key encryption protocol for quantum message. The encryption key $l$ of PQC has $2n$ bits, and the scheme in Section~\ref{subsec42} is used to encrypt single-bit classical message, so that the $2n$ bits should be encrypted one by one.

Alice intends to encrypt a $2n$-bit number $l$, thus she must get Bob's quantum public-keys. There are two requirements for the quantum public-keys: (1) In order to protect $k$, all the quantum public-keys are different (See the proof of Proposition 1); (2) In order to encrypt $2n$ bits, she needs to get $2n$ quantum public-keys (From the scheme in Section~\ref{subsec42}, Alice stores single-bit message in a quantum public-key, and then sends the quantum public-key to Bob. However, she does not know the state of the quantum public-key, and cannot produce its copies according to quantum no-cloning theorem. Thus, if she has only one copy of Bob's public-key, she can only encrypt one bit). From the two requirements, Alice must get $2n$ different quantum public-keys published by Bob. We denote the $2n$ quantum public-keys as $(s_j,\rho_{k_j,i_j}),j\in\{1,\ldots,2n\}$, and the $j$-th bit ($j\in\{1,\ldots,2n\}$) of $l$ is $l_j$. Alice encrypts each bit $l_j$, and obtains a quantum cipher $\rho_{k_j,i_j}^{(l_j)}=\mathcal{E}(\rho_{k_j,i_j},l_j)$.

If Alice intends to send a $n$-qubit message $\sigma$ to Bob securely, she firstly downloads $2n$ quantum public-keys: $(s_j,\rho_{k_j,i_j}),j\in\{1,\ldots,2n\}$ from Bob's public-key register.
Let $\rho_{ki}=(\rho_{k_1,i_1},\cdots,\rho_{k_{2n},i_{2n}})$. The encryption process is as follows.

{\bf [Encryption]}
\begin{enumerate}
\item{}
Alice randomly selects a $2n$-bit number $l=(l_1,\ldots,l_{2n})\in\{0,1\}^{2n}$;
\item{}
Alice encrypts the quantum message $\sigma$ with PQC encryption transformation and the classical key $l$, and obtains $l^{(\sigma)}=PQC_l(\sigma)$;
\item{}
Alice uses Bob's public-key $(s_j,\rho_{k_j,i_j})$ to encrypt each bit $l_j$($j\in\{1,\ldots,2n\}$), and obtains $2n^2$ qubits
$$\rho_{ki}^{(l)}=\mathcal{E}'(\rho_{ki},l)\equiv\otimes_{j=1}^{2n}\mathcal{E}(\rho_{k_j,i_j},l_j)\equiv\otimes_{j=1}^{2n}\rho_{k_j,i_j}^{(l_j)};$$
\item{}
Alice sends all the $2n$ strings $s_1,\ldots,s_{2n}$ and quantum cipher $\rho_{ki}^{(l)}\otimes l^{(\sigma)}$ to Bob.
\end{enumerate}

Bob receives these classical numbers $s_1,\ldots,s_{2n}$ and the quantum cipher $\rho_{ki}^{(l)}\otimes l^{(\sigma)}$, and then performs the decryption process.

{\bf [Decryption]}
\begin{enumerate}
\item{}
According to $s_1,\ldots,s_{2n}$, Bob uses his private-key $F$ to compute $k_j=F(s_j),j\in\{1,\ldots,2n\}$;
\item{}
Bob uses $k_1,\ldots,k_{2n}$ to decrypt the first $2n^2$-qubit of the quantum cipher. He decrypts $\otimes_{j=1}^{2n} \rho_{k_j,i_j}^{(l_j)}$ and obtains each bit of $l$: $l_j=\mathcal{D}(\rho_{k_j,i_j}^{(l_j)},k_j),j\in\{1,\cdots,2n\}$;
\item{}
According to $l$, Bob decrypts quantum cipher $l^{(\sigma)}$ by performing PQC decryption transformation, and obtains the quantum message $\sigma=PQC_l^{-1}(l^{(\sigma)})$.
\end{enumerate}

The attacker Eve does not know the random string $l$ and the quantum public-keys used by Alice. Thus, with respect to Eve, a $n$-qubit state $\sigma$ is encrypted into a $2n^2+n$-qubit state
\begin{eqnarray}
\mathcal{G}(\sigma)&=&\frac{1}{2^{2n}}\sum_{l\in\{0,1\}^{2n}}\left[\left(\frac{1}{2^{2n-1}}\right)^{2n}\sum_{\stackrel{k_1,\cdots,k_{2n}\in\Omega_n,}{i_1,\cdots,i_{2n}\in\{0,1\}^n}}\mathcal{E}'(\rho_{ki},l)\right]\otimes PQC_l(\sigma) \nonumber\\
&=&\frac{1}{2^{2n}}\sum_{l\in\{0,1\}^{2n}}\otimes_{j=1}^{2n}\left[\frac{1}{2^{2n-1}}\sum_{k_j\in\Omega_n}\sum_{i_j\in\{0,1\}^n}\rho_{k_j,i_j}^{(l_j)}\right]\otimes PQC_l(\sigma),
\end{eqnarray}
where $\mathcal{E}'(\rho_{ki},l)\equiv\otimes_{j=1}^{2n}\mathcal{E}(\rho_{k_j,i_j},l_j)$, $\rho_{ki}\equiv(\rho_{k_1,i_1},\cdots,\rho_{k_{2n},i_{2n}})$.

In the above scheme, PQC encryption needs $2n$-bit classical number $l$. If we consider to replace PQC with APQC (replace the transformation $PQC_l(\sigma)$ with $APQC_l(\sigma)$ in the above protocol), the length of $l$ can be reduced. This method can save quantum resources while encrypting classical string $l$.

For example, we consider the case that, the PQC in Figure~\ref{fig1} is replaced with the APQC which is described in Eq.(\ref{eqn6}). Then $APQC_l(\sigma)=APQC_{a,b}(\sigma)$, where $l=a||b$ is an $m$-bit string. Thus, encrypting a $n$-qubit message $\sigma$ can obtain a $nm+n$-qubit cipher. The cipher with respect to Eve is
\begin{eqnarray}
\mathcal{G}'(\sigma)&=&\frac{1}{2^m}\sum_{l\in\{0,1\}^m}\left[\left(\frac{1}{2^{2n-1}}\right)^m\sum_{\stackrel{k_1,\cdots,k_m\in\Omega_n,}{i_1,\cdots,i_m\in\{0,1\}^n}}\left[\otimes_{j=1}^m \mathcal{E}(\rho_{k_j,i_j},l_j)\right]\right]\otimes APQC_l(\sigma) \nonumber\\
&=&\frac{1}{2^m}\sum_{l\in\{0,1\}^m}\otimes_{j=1}^m\left[\frac{1}{2^{2n-1}}\sum_{k_j\in\Omega_n}\sum_{i_j\in\{0,1\}^n}\rho_{k_j,i_j}^{(l_j)}\right]\otimes APQC_l(\sigma).
\end{eqnarray}
From $m<2n$, it can be infered that $n(m+1)<2n^2+n$. That means the cipher of a $n$-qubit state is shortened.

\subsection{Security analysis}\label{subsec44}

The security of the quantum public-key encryption scheme proposed in Section~\ref{subsec43} is analyzed from two aspects: (1) the security of private-key; (2) the security of encryption.

Firstly, we consider the security of private-key. In this scheme, $k$ is an important number because it can be directly used for decryption. According to the Holevo theorem \cite{nielsen2000}, at most $n$-bit classical information can be obtained from a $n$-qubit public-key $$\rho_{k,i}=\frac{1}{2}(|i\rangle+|i\oplus k\rangle)(\langle i|+\langle i\oplus k|),\forall i\in\{0,1\}^n, k\in\Omega_n.$$
If Eve receives enough copies of a public-key $(s,\rho_{k,i})$, she can obtain the $n$-bit information of $k$, and then attack the communication between Alice and Bob. Thus, in order to protect the communication, the copies of each quantum public-key must be limited by an upper bound $\lambda$.
That means Bob publishes at most $\lambda$ copies of the quantum public-key $(s,\rho_{k,i})$ according to a pair of $(s,i)$, and then selects a new pair of $(s,i)$ to produce new quantum public-key. The following proposition proves that the upper bound $\lambda$ is equal to $1$.

\stepcounter{proposition}
{\bf Proposition \arabic{proposition}:} Given $c\geq 2$ copies of a quantum public-key $(s,\rho_{k,i})$, the value of $k$ can be extracted successfully with probability at least $\frac{1}{2}$.

{\bf Proof:}
For arbitrary quantum public-key, suppose the $n$-qubit state is $\rho_{k,i}=\frac{1}{2}(|i\rangle+|i\oplus k\rangle)(\langle i|+\langle i\oplus k|)$, where $i\in\{0,1\}^n$ and $k\in\Omega_n$. Suppose Eve has received sufficient copies of $\rho_{k,i}$.
Firstly, she measures the first copy of $\rho_{k,i}$ in the basis $\{|0\rangle,|1\rangle\}$, and gets a $n$-bit string $r_1$. Then she measures the second copy of $\rho_{k,i}$ and gets the second string $r_2$. If $r_1=r_2$, she continues to measure the $t$-th ($t=3,4,\cdots$) copy of $\rho_{k,i}$, until the $t$-th string $r_t\neq r_1$. At this time, she can conclude $k=r_1\oplus r_t$.

We denote random variable $N$ as the measurement times until $k$ being determined. The probability of the number $k$ being determined until the $t$-th measurement is $$Pr(N=t)=\frac{1}{2^{t-1}},t\geq 2.$$ Thus expected value of $N$ is $$\sum_{t=2}^{+\infty} t\cdot Pr(N=t)=\sum_{t=2}^{+\infty}\frac{t}{2^{t-1}}=3.$$ That means, measurement for three times in average can determine the value of $k$. Moreover, $Pr(N=2)=\frac{1}{2}$, which means the successful probability is $\frac{1}{2}$ when there are two copies of $\rho_{k,i}$. $\hfill{}\Box$

According to Proposition 1, in order to protect $k$, only one copy of each quantum public-key $(s,\rho_{k,i})$ is permitted to be produced from a pair of $(s,i)$. Therefore, any two quantum public-keys published by Bob are different.
The attacker Eve can only obtain one copy of $\rho_{k,i}$. When she measures it, she will get $i$ and $i\oplus k$ both with probability $\frac{1}{2}$,  but cannot get both the values of $i$ and $i\oplus k$. Extracting the value of $k$ from $i$ or $i\oplus k$ is the same as attacking one-time-pad in classical cryptography. Therefore, extracting the value of $k$ from only one copy of $(s,\rho_{k,i})$ is information-theoretically impossible. Moreover, extracting the relation (the private-key $F$) between $s$ and $k$ is also information-theoretically impossible. There maybe exist some different quantum public-keys corresponding to the same $k$, such as $\rho_{k,i_1},\ldots,\rho_{k,i_t}$. In this case, Theorem 6 in \cite{pan2010} has proved that it is still information-theoretic secure if $t=o(n)$.

In the scheme proposed in \cite{kawachi2008}, one private-key corresponds to only one quantum public-key, but one quantum public-key can be published by many copies. In order to prevent the attacker to extract private-key from sufficiently more copies of quantum public-key, the number of the published copies must be limited. This scheme can only be used to encrypt very limited number of qubits.
While in our scheme, one private-key corresponds to an exponential number of quantum public-keys, and any two quantum public-keys are different. Each quantum public-key can be used to encrypt one bit using the protocol proposed in Section~\ref{subsec42}, and two bits can be used to encrypt one qubit using PQC. Thus, the combined scheme can be used to encrypt an exponential number of qubits.

Next, we analyze the security from the second aspect. We prove the scheme in Section~\ref{subsec43} is information-theoretic secure.

\stepcounter{lemma}
{\bf Lemma \arabic{lemma}:} $D(\rho_1\otimes\rho_2,\rho_1'\otimes\rho_2')\leq D(\rho_1,\rho_1')+D(\rho_2,\rho_2')$, where $\rho_1,\rho_1',\rho_2,\rho_2'$ are any four density matrixes.

{\bf Proof:}
For any density matrix $\rho$, it is positive semidefinite and its trace is equal to 1. Thus, $tr|\rho|=tr(\rho)=1$.
\begin{eqnarray*}
&& D(\rho_1\otimes\rho_2,\rho_1'\otimes\rho_2')\\
&=&\frac{1}{2}tr|\rho_1\otimes\rho_2-\rho_1'\otimes\rho_2'|\\
&=&\frac{1}{2}tr|\rho_1\otimes\rho_2-\rho_1\otimes\rho_2'+\rho_1\otimes\rho_2'-\rho_1'\otimes\rho_2'|\\
&\leq &\frac{1}{2}tr|\rho_1\otimes(\rho_2-\rho_2')|+\frac{1}{2}tr|(\rho_1-\rho_1')\otimes\rho_2'|\\
&=&\frac{1}{2}tr|\rho_1|\cdot tr|\rho_2-\rho_2'|+\frac{1}{2}tr|\rho_1-\rho_1'|\cdot tr|\rho_2'|\\
&=&\frac{1}{2}tr|\rho_2-\rho_2'|+\frac{1}{2}tr|\rho_1-\rho_1'|\\
&=& D(\rho_1,\rho_1')+D(\rho_2,\rho_2').\hfill{~~~~~~~~~~~~~~~~~~~~~~~~~~~~~~~~~~~~~~~~~~~~~~~~~}\Box
\end{eqnarray*}

\stepcounter{proposition}
{\bf Proposition \arabic{proposition}:} There exists a fixed quantum state $\tau$, such that $D(\mathcal{G}(\sigma),\tau)\leq\frac{n}{2^{n-2}},\forall\sigma$.

{\bf Proof:} According to Lemma 4 in \cite{pan2010}, it can be inferred that $\forall j\in\{1,\cdots,2n\}$,
\begin{equation}
D\left(\frac{1}{2^{2n-1}}\sum_{k_j\in\Omega_n}\sum_{i_j\in\{0,1\}^n}\mathcal{E}(\rho_{k_j,i_j},l_j), \frac{1}{2^{2n-1}}\sum_{k_j\in\Omega_n}\sum_{i_j\in\{0,1\}^n}\mathcal{E}(\rho_{k_j,i_j},l_j\oplus 1)\right)=\frac{1}{2^{n-1}}.
\end{equation}

From Lemma 3, it can be known that $\forall l,l'\in\{0,1\}^{2n}$,
\begin{eqnarray}\label{eqn7}
&& D\left(\frac{1}{2^{2n(2n-1)}}\sum_{k,i}\mathcal{E}'(\rho_{ki},l),\frac{1}{2^{2n(2n-1)}}\sum_{k,i}\mathcal{E}'(\rho_{ki},l')\right)\nonumber\\
&\leq & 2n\cdot\frac{1}{2^{n-1}}=\frac{n}{2^{n-2}},
\end{eqnarray}
where the notation $\sum\limits_{k,i}$ represents $\sum\limits_{k_1,\cdots,k_{2n}}\sum\limits_{i_1,\cdots,i_{2n}}$ and $\rho_{ki}=(\rho_{k_1,i_1},\cdots,\rho_{k_{2n},i_{2n}})$.
From Lemma 1, $\exists \tau_1$, such that
\begin{equation}\label{eqn11}
D\left(\frac{1}{2^{2n(2n-1)}}\sum_{k,i}\mathcal{E}'(\rho_{ki},l),\tau_1\right)\leq\epsilon_1=\frac{n}{2^{n-2}},\forall l\in\{0,1\}^{2n}.
\end{equation}

According to the definition of PQC, there exists a fixed quantum state $\tau_2$, which satisfies
\begin{equation}
D\left(\frac{1}{2^{2n}}\sum_{l\in\{0,1\}^{2n}} PQC_l(\sigma),\tau_2\right)\leq\epsilon_2=0,\forall\sigma.
\end{equation}
Then, from Theorem 2, $\exists\tau=\tau_1\otimes\tau_2$, such that
\begin{equation*}
\hfill{~~~~~~~~~~~~~~~~~~~~~~~~}
D\left(\mathcal{G}(\sigma),\tau\right)\leq \epsilon_1+\epsilon_2=\frac{n}{2^{n-2}}.\hfill{~~~~~~~~~~~~~~~~~~~~~~~~}\Box
\end{equation*}

According to Proposition 2 and Corollary 1, the public-key encryption scheme proposed in Section~\ref{subsec43} is information-theoretic secure.

Finally, we study the security of the public-key encryption scheme when its PQC module is replaced with an APQC. Suppose APQC satisfies a condition: there exists a fixed quantum state $\tau_2$, such that for any positive polynomial $p(n)$, and all sufficiently large $n$, it holds that
\begin{equation}\label{eqn3}
D\left(\frac{1}{2^m}\sum_{l\in\{0,1\}^m}APQC_l(\sigma),\tau_2\right)\leq \epsilon_2=\frac{1}{p(n)},\forall\sigma.
\end{equation}
We will show the modified scheme is still information-theoretic secure.

\stepcounter{proposition}
{\bf Proposition \arabic{proposition}:} If the PQC is replaced with an APQC satisfying the condition Eq.(\ref{eqn3}), there exists a fixed quantum state $\tau$, such that for any positive polynomial $p(n)$, and all sufficiently large $n$, $D(\mathcal{G}'(\sigma),\tau)\leq \frac{1}{p(n)},\forall\sigma$.

{\bf Proof:}
This proof is similar to the proof of Proposition 2. Because the length of $l$ is $m<2n$ in the modified scheme, the right formula of Eq.(\ref{eqn7}) should be modified to be the expression "$\leq \frac{m}{2^{n-1}}<\frac{n}{2^{n-2}}$". Therefore,
\begin{equation}
D\left(\frac{1}{2^{m(2n-1)}}\sum_{k,i}\mathcal{E}'(\rho_{ki},l),\tau_1\right)<\epsilon_1=\frac{n}{2^{n-2}},\forall l\in\{0,1\}^{m}.
\end{equation}
Because APQC satisfies the condition Eq.(\ref{eqn3}) and according to Theorem 2, there exists a state $\tau=\tau_1\otimes\tau_2$, such that for any positive polynomial $p(n)$, and all sufficiently large $n$, it holds that $D(\mathcal{G}'(\sigma),\tau)<\epsilon_1+\epsilon_2=\frac{n}{2^{n-2}}+\frac{1}{p(n)}$. Therefore the proposition is proved.$\hfill{}\Box$

It can be inferred from Proposition 3 and Corollary 1 that, the modified scheme is still information-theoretic secure when APQC satisfies the condition Eq.(\ref{eqn3}).

\section{Public-key authentication of quantum information}\label{sec5}
\subsection{A general construction}\label{subsec51}
Barnum et al. \cite{barnum2002} proposed a symmetric-key authentication of quantum messages. In their scheme, Alice and Bob must preshare some classical bit-strings. We will construct an asymmetric-key authentication scheme for quantum messages, where the public-key is quantum state.

In a symmetric-key authentication of quantum messages, the authentication transformation is denoted as $u^{(\sigma)}=A_u(\sigma)$, where $\sigma$ is a quantum message, $u$ is a classical parameter for authentication.
See \cite{barnum2002} for a concrete instance of this type.

In a quantum public-key encryption of classical messages (for example $\Delta$ in Section~\ref{subsec41}), the encryption transformation is denoted as $\rho_k^{(u)}=\mathcal{E}(\rho_k,u)$, where $u$ is a classical bit-string to be encrypted, and $\rho_k$ is a quantum public-key.

A general public-key authentication of quantum messages is described as follows.
\begin{enumerate}
\item{}
Alice randomly selects a classical bit-string $u$, and then performs the authentication transformation on the quantum message $\sigma$, and obtains a quantum state $u^{(\sigma)}=A_u(\sigma)$;
\item{}
Alice uses Bob's quantum public-key $\rho_k$ to encrypt $u$, and obtains a quantum cipher $\rho_k^{(u)}=\mathcal{E}(\rho_k,u)$;
\item{}
Alice sends $\rho_k^{(u)}$ and $u^{(\sigma)}$ to Bob;
\item{}
Bob receives the quantum state $\rho_k^{(u)}\otimes u^{(\sigma)}$, and then decrypts $\rho_k^{(u)}$ using his private-key $F$, and obtains the classical string $u$;
\item{}
According to $u$, Bob authenticates $u^{(\sigma)}$ and recovers the message. If authentication succeeds, he can get the untampered quantum message $\sigma$.
\end{enumerate}

\subsection{Symmetric-key authentication of quantum information \cite{barnum2002}}\label{subsec52}

Barnum et al. \cite{barnum2002} defined authentication of quantum messages, and showed that any scheme to
authenticate quantum messages must also encrypt them. They also constructed a non-interactive symmetric-key authentication scheme with inherent encryption. In their scheme, a stabilizer purity testing code (SPTC) $\{Q_z\},z\in\mathcal{K}$ is used for authentication. Alice and Bob must preshare three random strings $z,x,y$.

Next, we describe their authentication scheme briefly. We denote $E_x(\sigma)$ as its encryption transformation which encrypts quantum message $\sigma$ with classical key $x$, and denote its decryption transformation as $D_x(*)$. The secret key $x$ is used in both encryption and decryption processes.
(For example, the PQC can be used here.)

Alice intends to send a quantum message $\sigma$ to Bob. The authentication process is as follows.
\begin{enumerate}
\item{}
Alice encrypts the quantum message $\sigma$ using the key $x$, and obtains $x^{(\sigma)}=E_x(\sigma)$;
\item{}
By using $z$ and $y$, Alice encodes $x^{(\sigma)}$ for the code $Q_z$ with syndrome $y$ to produce $u^{(\sigma)}$, where $u=x||z||y$; Alice sends $u^{(\sigma)}$ to Bob;
\item{}
Bob receives $u^{(\sigma')}$, and measures the syndrome $y'$ of the code $Q_z$ on this quantum state. Then he compares $y$ to $y'$, and aborts if $y\neq y'$; otherwise, it can be inferred that $u^{(\sigma')}=u^{(\sigma)}$, and the next step continues;
\item{}
Bob decodes $u^{(\sigma)}$ according to $Q_z$, and obtains $x^{(\sigma)}$. Bob decrypts $x^{(\sigma)}$ using $x$, and obtains the quantum message $\sigma=D_x(x^{(\sigma)})$.
\end{enumerate}

\subsection{Quantum public-key authentication of quantum information}\label{subsec53}
According to the general construction in Section~\ref{subsec51}, a public-key authentication scheme for quantum messages can be constructed by combining a quantum public-key encryption scheme $\Delta$ with the authentication scheme as in Section~\ref{subsec52}. Here, the scheme introduced in Section~\ref{subsec42} is selected as the quantum public-key encryption scheme $\Delta$, and the PQC is selected as the encryption module of the authentication scheme as in Section~\ref{subsec52}.

Next, we present the quantum public-key authentication scheme in detail. We denote $Q_{z,y}(\sigma)$ as encoding $\sigma$ for the code $Q_z$ with syndrome $y$. Let $u=x||z||y$, the authentication transformation is $A_u(\sigma)=A_{x,z,y}(\sigma)=Q_{z,y}(PQC_x(\sigma))$.

The SPTC $\{Q_z\}(z\in\mathcal{K})$ is public. Alice authenticates a $n$-qubit state $\sigma$ as follows.
\begin{enumerate}
\item{}
Alice randomly selects three parameters $x,z,y$, where $x\in\{0,1\}^{2n},z\in\mathcal{K}, y\in\mathcal{S}$, and $\mathcal{S}$ is a set of correctable syndromes. Suppose the total length of the three parameters is $h$;
\item{}
According to $u$ (or $x||z||y$), Alice performs authentication transformation on $\sigma$, and obtains $u^{(\sigma)}=A_u(\sigma)=Q_{z,y}(PQC_x(\sigma))$;
\item{}
Alice uses Bob's quantum public-keys $(s_j,\rho_{k_j,i_j}),j=1,\ldots,h$ to encrypt the classical bit-string $u\equiv(u_1,\ldots,u_{h})$, and obtains $\rho_{k_j,i_j}^{(u_j)}=\mathcal{E}(\rho_{k_j,i_j},u_j),j=1,\ldots,h$;
\item{}
Alice sends $s_1,\cdots,s_h$ and $\otimes_{j=1}^h\rho_{k_j,i_j}^{(u_j)}\otimes u^{(\sigma)}$ to Bob.
\end{enumerate}

Bob uses his private-key $F$ to authenticate the integrity of quantum messages as follows.
\begin{enumerate}
\item{}
Bob uses his private-key $F$ to decrypt $\rho_{k_j,i_j}^{(u_j)},j=1,\ldots,h$, and obtains $u$ (here $u=x||z||y$);
\item{}
According to the authentication process in Section~\ref{subsec52}, Bob uses $u$ to authenticate $u^{(\sigma)}$ (or $Q_{z,y}(PQC_x(\sigma))$).
\end{enumerate}

From Section~\ref{subsec42}, Bob's quantum public-key is $(s,\rho_{ki})$, where $s=(s_1,\cdots,s_h)$ and $\rho_{ki}=(\rho_{k_1,i_1},\ldots,\rho_{k_h,i_h})$. Alice uses Bob's public-key $(s,\rho_{ki})$ to encrypt the parameter $u=x||z||y$. The encryption transformation can be represented as
$\mathcal{E}'(\rho_{ki},u)=\otimes_{j=1}^h \mathcal{E}(\rho_{k_j,i_j},u_j)$. Thus, the public-key authentication transformation on the quantum message $\sigma$ using the quantum public-key $\rho_{ki}$ can be represented as
\begin{equation}
\mathcal{A}(\rho_{ki},\sigma)=\mathcal{E}'(\rho_{ki},x||z||y)\otimes Q_{z,y}\left(PQC_x(\sigma)\right),
\end{equation}
where $x\in\{0,1\}^{2n},z\in\mathcal{K},y\in\mathcal{S}$. The total length of the three parameters is $h=2n+log|\mathcal{K}|+log|\mathcal{S}|$.
Because the public-key and the random parameters $x||z||y$ are unknown to Eve, the state obtained after public-key authentication (with respect to Eve) is
\begin{equation}\label{eqn8}
\mathcal{F}(\sigma)
= \sum_{x,z,y}p_{x,z,y}\left(\frac{1}{2^{h(2n-1)}}\sum_{k,i}\mathcal{E}'(\rho_{ki},x||z||y)\otimes Q_{z,y}\left(PQC_x(\sigma)\right)\right),
\end{equation}
where $p_{x,z,y}=p_xp_yp_z$ is the probability of selecting $x||z||y$, and the encryption transformation $\mathcal{E}'$ is the same as introduced in Section~\ref{subsec43}. The notation $\sum\limits_{k,i}$ represents $\sum\limits_{k_1,\cdots,k_h}\sum\limits_{i_1,\cdots,i_h}$.

If the PQC in the above scheme is replaced with APQC, the public-key authentication transformation on the quantum message $\sigma$ using the quantum public-key $\rho_{ki}$ is represented as
\begin{equation}
\mathcal{A}'(\rho_{ki},\sigma)=\mathcal{E}'(\rho_{ki},x||z||y)\otimes Q_{z,y}\left(APQC_x(\sigma)\right),
\end{equation}
where $x\in\{0,1\}^m,z\in\mathcal{K},y\in\mathcal{S}$. The total length of the three parameters is $h'=m+log|\mathcal{K}|+log|\mathcal{S}|$.
With respect to Eve, the state obtained after public-key authentication is
\begin{equation}\label{eqn9}
\mathcal{F}'(\sigma)
= \sum_{x,z,y}p_{x,z,y}\left(\frac{1}{2^{h'(2n-1)}}\sum_{k,i}\mathcal{E}'(\rho_{ki},x||z||y)\otimes Q_{z,y}\left(APQC_x(\sigma)\right)\right),
\end{equation}
where $\sum\limits_{k,i}$ represents $\sum\limits_{k_1,\cdots,k_{h'}}\sum\limits_{i_1,\cdots,i_{h'}}$. The only difference between Eq.(\ref{eqn9}) and Eq.(\ref{eqn8}) is that $PQC_x(\sigma)$ has been replaced with $APQC_x(\sigma)$ and the total length of parameters $x||z||y$ decreases to $h'$ from $h$.

\subsection{Security analysis}
Barnum et al. \cite{barnum2002} showed that any scheme to
authenticate quantum messages must also encrypt them. The public-key authentication scheme proposed in Section~\ref{subsec53} has inherent encryption. Thus, the security must be considered from two aspects: (1) analyzing the security of authentication; (2) analyzing the security of encryption. In this authentication scheme, the authentication module is just the symmetric-key authentication scheme proposed in \cite{barnum2002}, which has the
error probability (or the success probability of tampering) decreasing exponentially with the security parameter (See Theorem 4 in \cite{barnum2002}). Therefore, while our scheme being used for authentication, the error probability decreases exponentially with the security parameter. Thus, we only need to consider its security with regard to encryption.

\stepcounter{proposition}
{\bf Proposition \arabic{proposition}:} There exists a fixed quantum state $\tau$, such that \\ $D(\mathcal{F}(\sigma),\tau)\leq\frac{h}{2^{n-1}},\forall\sigma$, where $h=2n+log|\mathcal{K}|+log|\mathcal{S}|$.

{\bf Proof:}
$\forall \sigma$
\begin{eqnarray}
\sum_u p_u A_u (\sigma)&=& \sum_{x,z,y}p_{x,z,y}Q_{z,y}(PQC_x(\sigma)) \nonumber\\
&=&\sum_{z,y}p_z p_y Q_{z,y}\left(\sum_x p_x PQC_x(\sigma)\right) \nonumber\\
&=& \sum_{z,y}p_z p_y Q_{z,y}(\sigma_0),
\end{eqnarray}
where $\sigma_0$ is a fixed state.
Thus, there exists a fixed quantum state $\tau_2=\sum_{z,y}p_z p_y Q_{z,y}(\sigma_0)$, such that
\begin{equation}
D(\sum_u p_u A_u (\sigma),\tau_2)\leq\epsilon_2=0.
\end{equation}

According to the way to deduce Eq.(\ref{eqn11}), there exists a fixed state $\tau_1$, such that
\begin{equation}\label{eqn10}
D\left(\frac{1}{2^{h(2n-1)}}\sum_{k,i}\mathcal{E}'(\rho_{ki},x||z||y),\tau_1\right)\leq \epsilon_1=\frac{h}{2^{n-1}},\forall x,z,y,
\end{equation}
where $h$ is the total length of $x,z,y$. According to Theorem 2, there exists a fixed state $\tau=\tau_1\otimes\tau_2$, such that $D(\mathcal{F}(\sigma),\tau)\leq\epsilon_1+\epsilon_2=\frac{h}{2^{n-1}}$.$\hfill{}\Box$

It can be inferred from Proposition 4 and Corollary 1 that, the public-key authentication scheme is information-theoretic secure with regard to encryption.

\stepcounter{remark}
{\bf Remark \arabic{remark}:} The public-key authentication scheme proposed in Section~\ref{subsec53} is analyzed with regard to encryption. This scheme can be seen as a combined construction of an encryption transformation $\mathcal{E}'$ for the classical string $u=x||z||y$ and an encryption (or authentication) transformation $A_u(*)$ for quantum message $\sigma$, so the result of Theorem 2 can also be used while analyzing its security with regard to encryption.

Next, we consider the security of modified public-key authentication scheme, where PQC is replaced with APQC.

\stepcounter{proposition}
{\bf Proposition \arabic{proposition}:} Suppose APQC satisfies the condition: there exists a fixed quantum state $\sigma_0$, such that for any $n$-qubit state $\sigma$, $$D\left(\sum_{x\in\{0,1\}^m} p_x APQC_x(\sigma),\sigma_0\right)\leq\epsilon.$$ Then, there exists a fixed state $\sigma_1$, such that
\begin{equation}
D\left(\sum_u p_u A_u'(\sigma),\sigma_1\right)\leq\epsilon,\forall\sigma,
\end{equation}
where $A_u'(\sigma)=Q_{z,y}(APQC_x(\sigma))$, $u=x||z||y$.

{\bf Proof:} The encoding transformation $Q_{z,y}(\cdot)$ is linear, so
\begin{eqnarray}
\sum_u p_u A_u'(\sigma)&=& \sum_{x,z,y}p_{x,z,y}Q_{z,y}(APQC_x(\sigma)) \nonumber\\
&=& \sum_{z,y}p_zp_yQ_{z,y}\left(\sum_x p_x APQC_x(\sigma)\right).
\end{eqnarray}
$Q_{z,y}(\cdot)$ is a quantum transformation for encoding, and does not change the trace distance of any two states. Therefore, there exists a fixed state $\sigma_1=\sum_{z,y}p_zp_yQ_{z,y}(\sigma_0)$, such that
\begin{eqnarray*}
&& D\left(\sum_u p_u A_u'(\sigma),\sigma_1\right)\nonumber\\
&\leq & \sum_{z,y}p_zp_yD\left(Q_{z,y}\left(\sum_x p_x APQC_x(\sigma)\right), Q_{z,y}\left(\sigma_0\right)\right)\nonumber\\
&=& \sum_{z,y}p_zp_yD\left(\sum_x p_x APQC_x(\sigma), \sigma_0\right)\nonumber\\
&\leq &\epsilon.\hfill{~~~~~~~~~~~~~~~~~~~~~~~~~~~~~~~~~~~~~~~~~~~~~~~~~~~~~~~~~~~~~~~~~~~~~~~~~~~~~~~~~}\Box
\end{eqnarray*}

Suppose APQC satisfies a condition: there exists a fixed quantum state $\sigma_0$, such that for every positive polynomial $p(n)$ and all sufficiently large $n$,
\begin{equation}\label{eqn4}
D\left(\sum_x p_x APQC_x(\sigma),\sigma_0\right)\leq \frac{1}{p(n)},\forall\sigma.
\end{equation}

\stepcounter{proposition}
{\bf Proposition \arabic{proposition}:} Suppose an APQC satisfying the condition Eq.(\ref{eqn4}) is used in the modified public-key authentication scheme, then there exists a fixed quantum state $\tau$, such that for every positive polynomial $p'(n)$ and all sufficiently large $n$, it has $D(\mathcal{F}'(\sigma),\tau)\leq\frac{1}{p'(n)},\forall\sigma$.

{\bf Proof:} Because the APQC satisfies the condition Eq.(\ref{eqn4}), it can be inferred from Proposition 5 that there exists a fixed state $\tau_2$, such that for every positive polynomial $p(n)$ and all sufficiently large $n$, then it holds that
\begin{equation}
D\left(\sum_u p_u A_u'(\sigma),\tau_2\right)\leq\epsilon_2=\frac{1}{p(n)},\forall\sigma.
\end{equation}
Similar to the deduction of Eq.(\ref{eqn10}), there exists a state $\tau_1$, such that
\begin{equation*}
D\left(\frac{1}{2^{h'(2n-1)}}\sum_{k,i}\mathcal{E}'(\rho_{ki},x||z||y),\tau_1\right)\leq \epsilon_1=\frac{h'}{2^{n-1}},\forall x,z,y.
\end{equation*}
According to Theorem 2, there exists a state $\tau=\tau_1\otimes\tau_2$, such that for every positive polynomial $p'(n)$ and all sufficiently large $n$, then it holds that
\begin{equation*}
D(\mathcal{F}'(\sigma),\tau)\leq\epsilon_1+\epsilon_2=\frac{h'}{2^{n-1}}+\frac{1}{p(n)}\leq\frac{1}{p'(n)},\forall\sigma.\hfill{~~~~~~~~~~~~~~}\Box
\end{equation*}

According to Proposition 6 and Corollary 1, if the APQC satisfies the condition Eq.(\ref{eqn4}), the modified public-key authentication scheme is still information-theoretic secure with regard to encryption.

\section{Discussion}
The public-key encryption scheme proposed in Section~\ref{subsec41} is different from classical public-key encryption. Here, the public-keys are quantum states and are unknown to the sender Alice. Alice uses the quantum public-keys without knowing their concrete states.
It is also different from the public-key encryption scheme proposed in \cite{kawachi2008}, because one private-key corresponds to an exponential number of quantum public-keys here. However, in \cite{kawachi2008} one private-key corresponds to only one public-key.

The quantum public-key encryption scheme suggested by Kawachi and Portmann \cite{kawachi2008} is bounded information-theoretic secure. However, our scheme is truly information-theoretic secure. The reason is as follows. In \cite{kawachi2008}, each quantum public-key can be used $O(n)$ times (strictly, each quantum public-key can have $O(n)$ copies, and each copy can be used one time), and can encrypt $n$ qubits each time, so each quantum public-key can be used to encrypt only
$O(n^2)$ qubits. Moreover, one private-key corresponds to only one public-key in the scheme, so each private-key can only be used to protect $O(n^2)$ qubits. Therefore, the scheme is bounded information-theoretic secure.
In our quantum public-key encryption scheme, a function $F$ is used as private-key. An exponential-size set generated from $F$ is denoted as
$\mathcal{R}=\{(k,i)|i\in\{0,1\}^n,k=F(s)\in\Omega_n,s\in\{0,1\}^{O(n)}\}$, where $|\mathcal{R}|=O(2^n)$. From each element $(k,i)$ of $\mathcal{R}$, a quantum state $\rho_{k,i}$ can be produced. Thus, one private-key $F$ corresponds to $O(2^n)$ quantum public-keys. Moreover, each quantum public-key can be used only once, and can encrypt $O(1)$ qubits, so each private-key $F$ can be used to protect $O(2^n)$ qubits. Therefore, our scheme is truly information-theoretic secure.

In the general construction as shown in Figure~\ref{fig1}, the classical number $l$ is encrypted using a type of quantum public-key scheme $\Delta$, then Alice and Bob can securely communicate without a preshared secret key. In order to construct an encryption scheme of quantum message without a preshared key,
there are another two choices to replace the module $\Delta$: (1) quantum secure direct communication (QSDC)\cite{long2002,deng2003,deng2004,zhu2006,gu2009,qin2009}; (2) quantum asymmetric cryptosystem proposed in \cite{gao2009}. The security needs further analysis for these choices.

Barnum et al. \cite{barnum2002} proposed a public-key authentication scheme for quantum messages. In the scheme, the sender uses classical cryptosystem to encrypt and sign the parameters $x,z,y$, and obtains classical cipher, and then sends it and the authenticated state of $\sigma$.
Because all the classical public-key cryptosystoms are based on computationally hard problems and are computational secure, thus the authentication scheme is computational secure.
However, in the quantum public-key authentication scheme in Section~\ref{subsec53}, the classical parameters $x,z,y$ are encrypted using a quantum public-key encryption scheme, which is information-theoretic secure.

Since lacking of digital signature, our authentication scheme cannot prevent Eve from substituting the quantum messages sent by Alice. It can only prevent quantum messages from being tampered. Thus, the scheme can protect the integrity of quantum messages.

In both the public-key encryption and authentication scheme, it is always assumed that the quantum public-keys are distributed securely. This paper concentrates on the study of the algorithm of quantum public-key cryptosystems, and does not study the management of quantum public-keys. However, the management of quantum public-keys is a critical problem which needs to be addressed, such as quantum public key infrastructure (QPKI).

\section{Conclusions}

This paper studies public-key cryptosystems using quantum public-keys. Firstly, we define information-theoretic security for public-key encryption of quantum message, and give two sufficient conditions for information-theoretic security. Secondly, we define a model of quantum public-key encryption of the classical messages. Based on this, a general construction is proposed for quantum public-key encryption of quantum message. Then a concrete example is presented and is proved to be information-theoretic secure. Finally, we propose a general construction for public-key authentication, which can protect the integrity of the quantum messages. We also suggests a concrete example of quantum public-key authentication. The example is proved to be information-theoretic secure with regard to encryption, and the success probability of tampering decreases exponentially with the security parameter with regard to authentication.

\section*{Acknowledgement}
This work was supported by the National Natural Science Foundation of China (Grant No. 61173157).

\end{document}